\documentclass[12pt,tightenlines,eqsecnum,floats,aps,amsmath,amssymb,nofootinbib
,superscriptaddress,showpacs]{revtex4}
   
\usepackage{amsmath,amsthm,amssymb,amsfonts}

\usepackage{colordvi}

\numberwithin{equation}{section}
\numberwithin{thr}{section}
\numberwithin{chr}{section}
\numberwithin{df}{section}

\begin{document}
%%%%%%%%%%%%%%%%%%%%%%%%%%%%%%%%%%

\title{The polymer quantization in LQG: massless scalar field}
\author{Marcin
\surname{Domaga{\l}a}}\email{marcin.domagala@fuw.edu.pl}
\affiliation{Instytut Fizyki Teoretycznej, Uniwersytet Warszawski,
ul. Ho{\.z}a 69, 00-681 Warszawa, Polska (Poland)}

\author{Micha{\l} 
\surname{Dziendzikowski}}\email{michal.dziendzikowski@fuw.edu.pl}
\affiliation{Instytut Fizyki Teoretycznej, Uniwersytet Warszawski,
ul. Ho{\.z}a 69, 00-681 Warszawa, Polska (Poland)}

\author{Jerzy \surname{Lewandowski}}\email{jerzy.lewandowski@fuw.edu.pl}
\affiliation{Instytut
Fizyki Teoretycznej, Uniwersytet Warszawski,
ul. Ho{\.z}a 69, 00-681 Warszawa, Polska (Poland)}\affiliation{Institute for
Quantum Gravity (IQG), FAU Erlangen -- N\"urnberg, Staudtstr. 7, 91058
Erlangen, Germany}

\pacs{4.60.Pp; 04.60.-m; 03.65.Ta; 04.62.+v}
\begin{abstract} The polymer quantization of matter fields is a diffeomorphism
invariant framework compatible  with Loop Quantum Gravity. Whereas studied by
itself, it is not explicitly used in the known completely quantizable models of
matter coupled to LQG. In the current paper we apply the polymer quantization to
the model of massless scalar field coupled to LQG. We show that the polymer
Hilbert space of the field degrees of freedom times the LQG Hilbert space of the
geometry degrees of freedom admit the quantum  constraints of GR and accommodate
their explicit solutions. In this way the quantization can be completed. That
explicit way of solving the quantum constraints suggests interesting new ideas.
 
\end{abstract}

\maketitle

\section{Introduction}

\subsection{Our goal}
 A successful quantization of the gravitational field does not complete the
standard model of fundamental interactions. All the standard matter fields need
to be quantized in a compatible way. In particular, the standard Fock space
quantization is not available. In Loop Quantum Gravity
\cite{book1,book2,book3,review} a new diffeomorphism
invariant framework for quantum matter field operators was introduced. In
particular, the scalar field is quantized according to the polymer  quantization
\cite{AAJL,AAHSJL, TT1, WKJL1,WKJL2}. On the other hand,  more recent  quantum
models
of matter interacting with the quantum geometry of LQG seem not to need any
specific quantization of a scalar field itself
\cite{RovelliSmolin,BrownKuchar,GieselThiemanndust,DGKL,Pullin}. For example,
when the
scalar constraint of General Relativity is solved classically, it  swallows one
scalar field which effectively becomes a parameter labeling the observables.
Therefore, this scalar field is treated in a different way, than  other fields.
Another insight comes from the Loop 
Quantum Cosmology. Within that framework, whereas the homogeneous gravitational
degrees of freedom are polymer quantized, the homogeneous scalar field is
quantized in a standard Quantum Mechanics fashion. Hence, the framework is
inconsistent in the way the scalar field is quantized as opposed to the
gravitational field. A third example is the full LQG model of the massless
scalar field coupled to gravity \cite{DGKL}. The final formulation of the model
is exact and precise, the Hilbert space and the quantum physical hamiltonian are
clearly defined modulo the issue of the self-adjoint extensions which is not
addressed. However, the derivation that leads to that result assumes the
existence of a suitable quantization of the scalar which is not used
explicitly.   
 
The goal of our current work is to show, that the polymer quantization of matter
fields can be used for coupling them with LQG. We demonstrate it on two known
examples of massless scalar field coupled to gravity: (i) a warming up example
is the homogeneous isotropic model of Loop Quantum Cosmology \cite{LQC1,LQC2},
and (ii)
the main  example is the case with all  the local degrees of freedom of the full
Loop Quantum Gravity \cite{DGKL}.            

\subsection{The Polymer quantization}  \label{poly_rep}
We recall here the Polymer quantization. Consider an $n$-dimensional real
manifold $\Sigma$ (a $3$D Cauchy surface in the case of GR), a real valued
scalar field $\varphi:\Sigma\rightarrow \mathbb{R}$,  the canonically conjugate
momentum $\pi$, and the Poisson bracket
\begin{equation}
\{\varphi(x),\pi(y)\}\ =\ \delta(x,y), \ \ \  \{\varphi(x),\varphi(y)\}\ =\ 0\ =
\ \{\pi(x),\pi(y)\}. 
\end{equation} 
Notice, that $\pi$ is a density of weight $1$, that is, upon a change of
coordinates $(x^a)\mapsto (x'^a)$ it transforms as a measure, that is 
\begin{equation}
\pi(x) d^nx \ =\ \pi(x') d^nx' .
\end{equation}
A Polymer variable representing $\pi$ is defined for every open, finite
${\cal{U}} \subset \Sigma$,  
\begin{equation}
\pi(V)\ =\ \int_Vd^nx\pi(x).
\end{equation}
A Polymer variable $U_p$ representing $\varphi$ is assigned to every function 
$p:\Sigma\rightarrow \mathbb{R}$
$$x\ \mapsto p_x$$
 of a {\it finite} support,
\begin{equation}
U_p(\varphi)\ =\ e^{i\sum_{x\in \Sigma}p_x\varphi(x)}.
\end{equation}
In particular  
\begin{equation}
U_{p=0}(\varphi)\ =\ 1.
\end{equation}
Notice, that 
\begin{equation}
{\rm supp}\,p\ =\ \{x_1,...,x_n\}, \ \ \ U_p(\varphi)\ =\
e^{i(p_{x_1}\varphi(x_1)+...+p_{x_n}\varphi(x_n))}
\end{equation}
The Poisson bracket between the Polymer variables is  
\begin{equation}
\{U_\pi, \pi(V)\}\ =\ i\left(\sum_{x\in V}p_x\right)\,U_p, \ \ \ \ 
\{U_p,U_{p'}\}\ =\ 0\ = \ \{\pi(V) , \pi(V')\}. 
\end{equation}
The Polymer quantization consists in using the following vector space 
\begin{equation}\label{Hmat}
{\{a_1U_{p_1}\ +\ ...\ + a_kU_{p_k}\ :\ a_I\in \mathbb{C}, k\in\mathbb{N}\}}
\end{equation}
endowed with the following Hilbert product
\begin{equation}
(\ U_p\ |\ U_{p'}\ )\ \ =\ \delta_{p,p'},
\end{equation} 
where the Kronecker delta takes values $0$ or $1$. That is we introduce the
Hilbert space
\begin{equation} {\cal H}\ :=\ \overline{\{a_1U_{p_1}\ +\ ...\ + a_kU_{p_k}\ :\
a_I\in \mathbb{C}, k\in\mathbb{N}\}}.\label{Hpoly} \end{equation}
Considered as an element of ${\cal H}$, the function $U_p$ will be denoted
by
\begin{equation}
U_p\ =:\ |p\rangle.
\end{equation}
The Polymer variables give rise to the Polymer operators
\begin{equation}
\hat{U}_p |p'\rangle\ =\ |p+p'\rangle, \ \ \ \hat{\pi}(V)|p\rangle\ =\ \hbar
\left(\sum_{x\in V}p_x\right)\, \ |p\rangle 
\end{equation}
Hence, the values $p_x$ taken by the function $p$ account to the spectrum of the
$\hat{\pi}(V)$ operators. For this reason, in the quantum context we will modify
the notation and write
$$ \hat{\pi}(V)|\pi\rangle\ =\ \hbar(\sum_{x\in\Sigma} \pi_x)|\pi\rangle, \ \ \
\ \ \ \ 
\hat{U}_\pi |\pi'\rangle\ =\ |\pi+\pi'\rangle $$
denoting by $\pi$ and $\pi'$  functions of the compact support
$$ x\mapsto \pi_x, \pi'_x\in \mathbb{R}. $$ 

The advantage of the polymer quantization is that the diffeomorphism of $\Sigma$
act naturally as unitary operators in the Hilbert space. This is what makes this
quantization different from the standard one.  

{\bf Remark}  {\it Diffeomorphism  invariant quantizations of the Polymer
variables
were studied in \cite{WKJL1,WKJL2} and a  class of inequivalent
quantizations parametrized by  a real parameter $a$ was found:}
\begin{equation}
\hat{\pi}(V)|\pi\rangle\ =\ \hbar \left(\sum_{x\in V}\pi_x\ +\ a{\rm E}(V)
\right)\, \ |\pi\rangle 
\end{equation}          
{\it where ${\rm E}(V)$ is the Euler characteristics of $V$
and $\hat{U}_\pi$ is the same as above,  independently on the value of $a$. 
However, nobody has ever used any of them for $a\not= 0$.} 
 
There is also a 1-degree of freedom ``poor man''  version  of the Polymer 
quantization that can be applied  to mechanics. 
Consider a variable $\Phi\in\mathbb{R}$ and the conjugate momentum $\Pi$,
and the Poisson bracket defined by
$$\{\Phi,\Pi\}\ =\ 1 \ \ \ \ \ \{\Phi,\Phi\}\ =\ 0\ =\ \{\Pi,\Pi\}. $$
The Polymer variables are $\Pi$ itself, and for every $p\in\mathbb{R}$, 
\begin{equation}
\tilde{U}_\pi(\Phi)\ :=\ e^{ip \Phi}.
\end{equation}
The Polymer quantum representation of those variables is defined 
in the seemingly usual way
\begin{equation}
\hat{U}_\pi\psi(\Phi)\ =\ \tilde{U}_\pi(\Phi)\psi(\Phi),\ \ \ \
\hat{\Pi}\psi(\Phi)\ =\ \frac{\hbar}{i}\frac{d}{d\Phi}\psi(\Phi),
\end{equation}
in an unusual  Hilbert space, though
\begin{equation}
\tilde{\cal H}\ :=\ \overline{\{a_1\tilde{U}_{\pi_1}\ +\ ...\ +
a_k\tilde{U}_{\pi_k}\ :\ a_I\in \mathbb{C}, k\in\mathbb{N},
\pi_I\in\mathbb{R}\}}
\end{equation}
with the Hilbert product defined such that the $U_\pi$ functions are
normalizable  
\begin{equation}
(\ \tilde{U}_\pi\ |\ \tilde{U}_{\pi'}\ )\ \ =\ \delta_{\pi,\pi'}.
\end{equation} 
If we again denote     
\begin{equation}
\tilde{U}_\pi\ =:\ |\pi\rangle,
\end{equation}
whenever it is considered an element of $\tilde{\cal H}$, then
\begin{equation} \label{poor_poly_rep}
\hat{\tilde{U}}_\pi|\pi'\rangle\ =\ |\pi+\pi'\rangle, \ \ \ \
\hat{\Pi}|\pi\rangle\ =\ \hbar \pi|\pi\rangle.
\end{equation}
Actually, even a polymer quantum mechanics was considered in the literature 
\cite{shadow1, shadow2,Fredenhagen}.       

The polymer quantization Hilbert spaces ${\cal H}$ and, respectively,
$\tilde{\cal H}$ can be obtained by suitable integrals. The poor man Hilbert
product can be defined by the Bohr measure such that
$$ \int_{\mathbb{\bar{R}_{\rm Bohr}}} d\mu_{\rm Bohr}(\Phi)e^{i\pi \Phi}\ =\
\delta_{0,\pi} $$  
where $\mathbb{\bar{R}_{\rm Bohr}}$ stands for the Bohr compactification of the
line. With certain abuse of notation we often write 
$$\tilde{\cal H}=L_2(\mathbb{\bar{R}}_{\rm Bohr}).$$ 
In the scalar field case, the polymer  Hilbert product is defined by the
infinite tensor product of the Bohr measures, that is the natural Haar
measure defined on the group $\mathbb{\bar{R}}_{\rm Bohr}{}^\Sigma$ of all the
maps $\Sigma\rightarrow \mathbb{\bar{R}}_{\rm Bohr}$. So one can write 
$${\cal H}\ =\ L_2(\mathbb{\bar{R}}_{\rm Bohr}{}^\Sigma).$$

\section{A doubly Polymer Quantization of LQC.}
A homogeneous and isotropic spacetime  coupled  to a KG scalar field
is described by two real valued dynamical variables $c,\Phi$, and their 
conjugate momenta $p_c, \Pi$. The Poisson bracket  $\{\cdot,\cdot\}$ is defined 
by 
\begin{equation} \{\Phi,\Pi\}\ =\ 1 \ = \{c,p_c\},\end{equation} 
whereas the remaining brackets vanish. 
The first  variable, $\Phi$, is the scalar field
constant on the homogeneous 3-manifold $\Sigma$.  The canonically conjugate
variable $\Pi$ is defined by a suitable integral of the momentum $\pi$, also
constant on $\Sigma$ by the homogeneity assumption. 
The variable $p$ is proportional to  the square of the scale of the universe
($a^2$), and $c$ to the rate of change in time ($\dot{a}$).

The constraints of General Relativity reduce to a single constraint, the Scalar
Constraint - and the Hamiltonian of the system - which for a massless scalar
field takes the following form \cite{APS}
\begin{equation}
C_{\pm} \ =\ \Pi\ \mp\ h(c,p_c),
\end{equation}
where $h$ is by definition a positive definite expression  (this is the
reduction of  the familiar $\sqrt{-2\sqrt{{\rm det q}}C_{\rm gr}}$ to the
homogeneous isotropic gravitational fields).

According to historically the first Wheeler de Witt quantization  of this model,
the both degrees of freedom are quantized in the usual way, that is the  
Hilbert space of the kinematical quantum states of the model is
$$ L_2(\mathbb{R})\otimes L_2(\mathbb{R}). $$

The LQC quantization uses the holonomy variables of Loop Quantum Gravity
restricted to the homogeneous isotropic solutions. The consequence is that the
gravitational degree of freedom $c$ ends up quantized in the Polymer way
\cite{AAMBJL}. The
scalar field, on the other hand, is quantized in the usual way. Finally, the
resulting Hilbert space of the kinematical quantum states of LQC is the hybrid
Hilbert space
$${L_2(\mathbb{R})\otimes L_2(\mathbb{\bar{R}}_{\rm Bohr})}. $$ 
Those details were set in this way without deeper thinking, just because
it works. 

The goal of this section is to present a fully Polymer formulation of
this LQC model in which the both variables $c$ and $\Phi$ are quantized in 
the Polymer way in the kinematical Hilbert space
\begin{equation}\label{Hkin}{\cal H}_{\rm kin}\ =\ L_2(\mathbb{\bar{R}}_{\rm
Bohr})\otimes
L_2(\mathbb{\bar{R}}_{\rm Bohr})\ =:\ {\cal H}_{\rm mat}\otimes {\cal H}_{\rm
gr}. \end{equation}
Let $\hat{\Pi}$ be the operator defined according to  \ref{poor_poly_rep}
in the first factor ${\cal H}_{\rm mat}$ polymer Hilbert space        
and let $\hat{h}$ be a quantum operator defined by a quantization
of the term $h(x,p)$ in the second factor ${\cal H}_{\rm gr}$ Polymer Hilbert
space.
Specifically, one can think of  the operator defined in \cite{APS},
or one of the wider class of operators considered in \cite{jlkamyk}.  
In fact, the operator is defined only in a suitable subspace of 
$${{\cal H}}_{{\rm gr},h}\subset\ {\cal H}_{\rm gr},$$
because it involves square roots of other operators which are not positive
definite,
and only the positive parts of their spectra are physical.
What will be important in this section is  that $\hat{h}$  is self-adjoint (it
is also non-negative) in  $\tilde{{\cal H}}_{{\rm gr},h}$.    
We will also have to reduce the full kinematical Hilbert space to
$$ \tilde{{\cal H}}_{{\rm kin},h}\ =\ {\cal H}_{\rm mat}\otimes {{\cal
H}}_{{\rm gr},h}.$$
The quantum constrain operator is
$$ \hat{C}_\pm\ =\ \hat{\Pi}\otimes {\rm id}\ \mp\ {\rm id}\otimes \hat{h}. $$
The Hilbert space   of the physical states is spanned by the two spaces 
$${\cal H}_{\rm phys\pm}\ =\ {\cal H}_{\hat{C}_{\pm}=0}$$
 of the spectral decompositions of the operators $\hat{C}_\pm$
 corresponding to $0$ in the spectrum of $\hat{C}_+$ (respectively $\hat{C}_-$).
 As we will see below, that   
space consists of  normalizable elements of (\ref{Hkin}).   

The main device we use is an operator $e^{i\hat{\Phi}\otimes \hat{h}}$. Itself,
an operator $\hat{\Phi}$ is not defined in the polymer Hilbert space, but the
definition of  $e^{i\hat{\Phi}\otimes \hat{h}}$ is quite natural if we use  
eigenvectors  $\{\psi_l\ :\  l\in L\}$ of the operator $\hat{h}$. 
 In ${\cal H}_{\rm kin}$ we consider the simultaneous eigenvectors of ${\rm
id}\otimes \hat{h}$ and $\hat{\Pi}\otimes {\rm id}$, that is
$$\{|\pi\rangle\otimes\psi_l\ :\  \pi\in \mathbb{R}, l\in L\}.$$ 
Define     
\begin{equation} \label{exphpi}
e^{i\hat{\Phi}\otimes \hat{h}} |\pi\rangle\otimes\psi_l\ :=\
e^{ih_l\hat{\Phi}}\otimes{\rm id} |\pi\rangle\otimes\psi_l\ =\
|\pi+h_l\rangle\otimes\psi_l. \end{equation}      
This operator preserves the norm, and admits inverse, namely
$$e^{-i\hat{\Phi}\otimes \hat{h}} |\pi\rangle\otimes\psi_l\ :=\
e^{-ih_l\hat{\Phi}}\otimes{\rm id} |\pi\rangle\otimes\psi_l\ =\
|\pi-h_l\rangle\otimes\psi_l. $$   
Therefore, it is a unitary operator in  ${{\cal H}}_{{\rm kin},h}$.

The next step in the derivation of the physical states, their Hilbert space,
and the Dirac observables is to notice that 
$$ \hat{C}_\pm\ =\ \hat{\Pi}\ \mp\  \hat{h}\ =\ e^{\pm
i\hat{\Phi}\otimes{\hat{h}}} \hat{\Pi} e^{\mp i\hat{\Phi}\otimes\hat{h}}. $$

For clarity, let us fix a sign in $\hat{C}_\pm$ and consider first, say,
$\hat{C}_+$. Indeed,  it follows that the spectrum decomposition of $\hat{C}_+$
is obtained from the spectral decomposition of $\hat{\Pi}$. In particular, the
Hilbert space corresponding to $0$ in the spectrum of $\hat{C}_+$ is obtained
from the Hilbert space of the decomposition of $\hat{\Pi}\otimes{\rm id}$ 
corresponding to $0$ in the spectrum, that is
$$ {\cal H}_{{\rm phys}+}\ =\ 
e^{i\hat{\Phi}\otimes{\hat{h}}}\left(|0\rangle\otimes {\cal H}_{{\rm
gr,h}}\right)\ \subset\ {\cal H}_{{\rm kin},h}.$$ 

Secondly, it follows that  
$$ [\hat{\cal O}, \hat{C}_+]\ =\ 0\ \Leftrightarrow
[e^{-i\hat{\Phi}\otimes{\hat{h}}}\hat{\cal O}
e^{i\hat{\Phi}\otimes\hat{h}},\hat{\Pi}\otimes{\rm id}]\ =\ 0.$$
The general solution for a Dirac observable is a function of the following basic
solutions 
$$ \hat{\cal O}^+_{\hat{L}}\ =\ e^{i\hat{\Phi}\otimes\hat{h}}{\rm
id}\otimes\hat{L} e^{-i\hat{\Phi}\otimes{\hat{h}}}, \ \ \ \ {\rm or}\ \ \ \
\hat{\cal O}\ =\ \hat{\Pi}\otimes {\rm id}.$$
The second option above, however, on ${\cal H}_{\rm phys+}$ reduces to 
\begin{equation}
\hat{\Pi}\otimes {\rm id}\ =\ {\rm id}\otimes \hat{h}\ =\ \hat{\cal
O}^+_{\hat{h}}.
\end{equation}

Next, we repeat the same construction for $\hat{C}_-$, derive ${\cal H}_{{\rm
phys},-}$, and the observables $\hat{\cal O}^-_{\hat{L}}$.  

The spaces ${\cal H}_{\rm phys}^\pm$ correspond to the non-negative/non-positive
eigenvalues of the scalar field momentum $\hat{\Pi}$. 
 They span a subspace
$${\cal H}_{\rm phys}\ \subset\ {\cal H}_{\rm kin}.$$ 
If $\hat{h}$ is bounded from zero (for example for negative cosmological
constant),
then 
$$ {\cal H}_{\rm phys}\ =\ {\cal H}_{{\rm phys}+}\ \oplus\ {\cal H}_{{\rm
phys}-}.$$
Otherwise, ${\cal H}_{{\rm phys}+}\cap {\cal H}_{{\rm phys}-}$ is the subspace 
of states $|\pi=0\rangle\otimes |h_l=0\rangle$. In both cases, the observables 
$\hat{\cal O}_{\hat{L}}^+$ and  $\hat{\cal O}_{\hat{L}}^-$ are consistent on the
overlap and give rise to observables defined on ${\cal H}_{\rm phys}$,
\begin{equation}
\hat{\cal O}_{\hat{L}}|_{{\cal H}_{{\rm phys},\pm}}\ =\  \hat{\cal
O}_{\hat{L}}^\pm. 
\end{equation}    

This result agrees with the known in the literature LQC model constructed
by the hybrid quantization, but it is quantized by applying consequently the
Polymer quantization to the both matter and gravity. This result generalizes in
the obvious way to the homogeneous non-isotropic models, because the Hilbert
space of the scalar field is unsensitive on that generalization.

\section{The polymer quantization of LQG} 
We turn now to the main subject of this work,  the scalar field  coupled to the
gravitational field. This section should be read as a continuation of the
lecture notes ,,From Classical To Quantum Gravity: Introduction to Loop Quantum
Gravity '' by Hanno Sahlmann and Kristina Giesel \cite{HannoKristina}, another
part of the current proceedings.         

The canonical field variables are defined on a 3-manifold $\Sigma$. They are
the scalar field $\varphi$ and its momentum $\pi$ introduced above in Section
\ref{poly_rep},
and the Ashtekar-Barbero variables $A^i_a$ and $E^j_b$.
     
The kinematical Hilbert space for the quantum scalar field (\ref{Hmat}) will be
denoted here by ${\cal H}_{\rm kin,mat}$. The kinematical Hilbert space for the
quantum gravitational field introduced in Section 3.1 of \cite{HannoKristina}
out of the cylindrical functions of the variable $A$ (connection),  
will be denoted here by ${\cal H}_{\rm kin,gr}$. The kinematical Hilbert space
for the system is
\begin{equation}
{\cal H}_{\rm kin}\ =\ {\cal H}_{\rm kin,mat}\otimes {\cal H}_{\rm kin,gr},
\end{equation}
and its elements are functions
$$(\varphi,A)\ \mapsto \psi(\varphi,A).$$

\subsection{The Yang-Mills gauge transformations and the Gauss constraint}
Classically, the theory is constrained by the the first class constraints:
the Gauss constraint, the vector constraint and the scalar constraint. 

The quantum Gauss constraint operator 
${\rm id}\otimes \hat{\cal G}(\Lambda)$, acts on the gravitational degrees of
freedom where the operator $\hat{\cal G}(\Lambda)$ is defined for every
$\Lambda:\Sigma\rightarrow \mathfrak{su}(2)$ in Section 3.2.1 of
\cite{HannoKristina}. The operator induces the unitary group of the ``Yang-Mills
gauge transformations'' acting in ${\cal H}_{\rm kin,gr}$,
\begin{equation}
\psi \mapsto\ U^{\rm G}(a)\psi, \ \ \ \ \ \ \ 
U^{\rm G}(a)\psi(\varphi,A)\ =\ \psi(\varphi,a^{-1}Aa+a^{-1}da). 
\end{equation}
The space of solutions to the Gauss constraint in ${\cal H}_{\rm kin,gr}$ 
was characterized at the end of Section 3.3.1 in \cite{HannoKristina} (and
denoted by ${\cal H}_{\rm kin}^G$). In the current paper, we will be denoting it
by ${\cal H}_{\rm kin,gr}^G$. 
The space of the solutions to the quantum Gauss constraint in ${\cal H}_{\rm
kin}$ is
\begin{equation} 
{\cal H}_{\rm kin}^G\ =\ {\cal H}_{\rm kin,mat}\otimes {\cal H}_{\rm kin,gr}^G. 
\end{equation} 
This is a subspace of ${\cal H}_{\rm kin}$ which consists of the elements
invariant
with respect to the Yang-Mills gauge transformations.
In terms of the generalized spin-networks, this subspace is the completion of
the span of the subspaces ${\cal H}_{\gamma,\vec{j},\vec{l}=0}$. 
There is an equivalent  constructive definition of the solutions called the
group averaging. It consists in integration with respect to the gauge
transformations
$$ \psi\ \mapsto\ \int \prod_{x\in \Sigma}da(x) \psi(\varphi,a^{-1}Aa+a^{-1}da)
$$
This kind of integral usually would be defined only ``formally''. However, if
$A\mapsto \psi(\varphi,A)$ is a function  cylindrical with respect to a graph
$\gamma$  embedded in $\Sigma$, then the Yang-Mills gauge transformations act
at the nodes $n_1,...,n_N$ of $\gamma$, in the sense that 
$$ \psi(\varphi,a^{-1}Aa+a^{-1}da)\ =\ f(\varphi,A,a(n_1),...,a(n_N)).$$
Therefore, 
\begin{equation}\label{Gaussaveraging}
\int \prod_{x\in \Sigma}da(x) \psi(\varphi,a^{-1}Aa+a^{-1}da) \ =\  \int
da(n_1)...da(n_N)\psi(\varphi,a^{-1}Aa+a^{-1}da)
\end{equation}
is actually defined very well.  This definition of solutions to the Gauss
constraint admits interesting generalization to the vector constraint.
 
Before solving the quantum vector constraint, we decompose the Hilbert space
suitably. To every finite set of points
\begin{equation}
X\ =\ \{x_1,...,x_k\}\subset \Sigma, 
\end{equation}
there is naturally assigned a subspace spanned by  the states $|\pi\rangle\in
{\cal H}_{\rm kin,mat}$ such that the support of the function
$\pi:\Sigma\rightarrow\mathbb{R}$ is exactly $X$
\begin{equation}
{\cal D}_{X}\ =\ {\rm Span}\left(\ |\pi\rangle\in {\cal H}_{\rm kin,mat}\ :\ 
{\rm supp}(\pi)=X\ \right). 
\end{equation}
The polymer Hilbert space ${\cal H}_{\rm kin,mat}$ is the completion of an
orthonormal sum of those subspaces
\begin{equation}
{\cal H}_{\rm kin,mat}\ =\ \overline{\bigoplus_{X\subset \Sigma\ :\
|X|<\infty}{\cal D}_X}.
\end{equation}
We will be precise about the domains of introduced maps, therefore we
distinguish
here explicitly between the span or infinite direct sum and the completion
thereof. 

The  Hilbert space ${\cal H}_{\rm kin,gr}^G$ of the gravitational degrees of
freedom
is also decomposed into orthogonal subspaces labeled by admissible graphs
embedded in $\Sigma$ (see the end of Section 3.3.1 of \cite{HannoKristina})
\begin{equation} 
{\cal H}_{\rm kin,gr}^G \ =\ \overline{\bigoplus_{\gamma}{\cal D}'_\gamma{}^G}
\end{equation}    
where $\gamma$ runs through the set of embedded graphs in $\Sigma$ 
admissible in the sense, that do not contain any 2-valent node that 
can be obtained splitting a single link and possibly reorienting the resulting
new links,  and   
\begin{equation}
{\cal D}'_\gamma{}^G\ =\ \bigoplus_{\vec{j}} {\cal H}_{\gamma,\vec{j},\vec{l}=0}
\end{equation}
where each $\vec{j}$ is a coloring of the links by irreducible non-trivial
representations of SU(2). The labeling $\vec{l}$, in general case, labels the
nodes of $\gamma$ by irreducible representations of SU(2), in this case it is
the trivial representation. The sum includes the empty graph $\emptyset$. A
cylindrical function with respect to the empty graph is a constant function.
        
The two decompositions are combined into the decomposition of the total Hilbert
space
\begin{equation}
{\cal H}_{\rm kin}^G\ =\ \overline{\bigoplus_{(X,\gamma)} {\cal D}_X\otimes{\cal
D}'_{\gamma}{}^G}\ .
\end{equation}
The uncompleted space
$$ {\cal D}_{\rm kin}^G\ :=\ \bigoplus_{(X,\gamma)} {\cal D}_X\otimes{\cal
D}'_{\gamma}{}^G, $$
will be   an important domain in what follows.  

\subsection{Diffeomorphisms and the vector constraint}
The diffeomorphisms of $\Sigma$ act naturally in ${\cal H}_{\rm kin}$,
\begin{equation}
{\rm Diff}\ni\phi\mapsto U(\phi)\in U({\cal H}_{\rm kin}),
\end{equation}
as the kinematical quantum states are functions of $A$ and $\phi$,     
\begin{equation}
U(\phi)\psi(\varphi,A)\ =\ \psi(\phi^*\varphi,\phi^*A).
\end{equation}
The only diffeomorphism invariant element of ${\cal H}_{\rm kin}$ is
$$\psi(\varphi,A)\ =\ {\rm const}.$$
However, the analogous to  (\ref{Gaussaveraging}) averaging with respect to the
diffeomorphisms  produces a larger than 1-dimensional Hilbert space, containing
also ``non-normalizable''  states. They become normalizable with respect to a
natural  Hilbert product. The diffeomorphism averaging in the matter free case
is discussed in detail in \cite{HannoKristina}. Now we need to discuss  it more
closely in the case with the scalar field. 

For each of the subspaces ${\cal D}_X\otimes{\cal D}'_{\gamma}{}^G$ introduced
above, denote by TDiff$_{X,\gamma}$ the set of the diffeomorphisms which act
trivially
in ${\cal D}_X\otimes{\cal D}'_{\gamma}{}^G$. It is easy to see that
TDiff$_{X,\gamma}$  consists of diffeomorphisms $\phi$ such that 
\begin{equation}
\phi|_X\ =\ {\rm id},\ \ \ {\rm and}\ \ \ \phi(\ell)=\ell \ \ {\rm for\ every}\
\  {\rm link}\ \ell\ {\rm of}\ \gamma,  
\end{equation}  
where we recall that the links are oriented, and the orientation has to be
preserved as well. We will average  with respect to the group of orbits 
\begin{equation}
{\rm Diff}/{\rm TDiff}_{X,\gamma}.
\end{equation} 
Given $\psi\in {\cal D}_X\otimes{\cal D}'_{\gamma}{}^G$, what is averaged is 
the dual state $\langle \psi|$, that is the linear functional on ${\cal H}_{\rm
kin}^{\rm G}$,  
$$\langle\psi|:\ \psi'\ \mapsto\ (\psi|\psi')_{\rm kin}.$$
The  averaging formula is simple: 
\begin{equation}
{\cal D}_X\otimes{\cal D}'_{\gamma}{}^G\ni \psi\ \mapsto\ \langle\psi|\ \mapsto\
\frac{1}{n_{X,\gamma}}\sum_{[\phi]\in {\rm Diff}/{\rm TDiff}_{X,\gamma}} \langle
U(\phi)\psi| \ =:\eta(\psi), 
\end{equation} 
where the factor $\frac{1}{n_{X,\gamma}}$ will be fixed below.
The result  of the averaging is a linear functional
$$ [\eta(\psi)](\psi')\ =\ \frac{1}{n_{X,\gamma}}\sum_{[\phi]\in {\rm Diff}/{\rm
TDiff}_{X,\gamma}} \left(U(\phi)\psi|\psi'\right)_{\rm kin}. $$
The map $\eta$  is defined for all the subspaces ${\cal D}_X\otimes{\cal
D}'_{\gamma}{}^G$ and extended by the linearity to their orthogonal sum ${\cal
D}_{\rm kin}^G$. 
As in the matter free case  \cite{HannoKristina}, one can also consider 
the subgroup Diff$_{X,\gamma}$ of the diffemorphisms preserving
the subspace ${\cal D}_X\otimes{\cal D}'_{\gamma}{}^G$. 
It gives rise to the symmetry group
\begin{equation}
{\rm GS}_{X,\gamma}\ :=\ {\rm Diff}_{X,\gamma}/{\rm TDiff}_{X,\gamma}
\end{equation}
which is finite. We fix the number $n_{X,\gamma}$ in the definition of $\eta$ to
be  
\begin{equation}
n_{X,\gamma}\ :=\ |{\rm GS}_{X,\gamma}|\ <\infty.
\end{equation}
The image of $\eta$  will be denoted as follows
$${\cal D}^G_{\rm Diff}\ :=\ \eta({\cal D}_{\rm kin}^G).$$  
The new scalar product in ${\cal D}^G_{\rm Diff}$ is
\begin{equation}
(\eta(\psi)\ |\ \eta(\psi'))_{\rm Diff}\ :=\ [\eta(\psi)](\psi'). 
\end{equation}
This completes the construction of the Hilbert space of the solutions
to the vector constraint 
\begin{equation}
{\cal H}^G_{\rm Diff}\ =\ \overline{\eta({\cal D}_{\rm kin}^G)}\ . 
\end{equation}
With this choice of $n_{X,\gamma}$, the map $\eta$  projects $\psi$ orthogonally
onto the subspace of   ${\cal D}_X\otimes{\cal D}'_{\gamma}{}^G$ consisting of
the elements symmetric with respect to the symmetries of $(X,\gamma)$, and next
unitarily embeds in ${\cal H}_{\rm Diff}^G$.
More generally,  for 
$$ \psi_I\in {\cal D}_{X_I}\otimes{\cal D}'_{\gamma_I}{}^G,\ \ \ \ I=1,2$$ 
the scalar product can be written in the following way
 \begin{equation}
(\eta(\psi_1)\ |\ \eta(\psi_2))_{\rm Diff}\ =\
\delta_{X_1,X_2}\delta_{\gamma_1,\gamma_2}(\psi_1|P_{X_2,\gamma_2}\psi_2)_{\rm
kin} 
\end{equation}
where  
$$P_{X_2,\gamma_2}\ :\  \overline{{\cal D}_{X_2}\otimes{\cal
D}'_{\gamma_2}{}^G}\rightarrow \overline{{\cal D}_{X_2}\otimes{\cal
D}'_{\gamma_2}{}^G}$$
is the orthogonal projection onto the subspace of states symmetric with respect
to the group GS$_{X_2,\gamma_2}$.    

The space ${\cal H}^G_{\rm Diff}$ is our Hilbert space of solutions
to the quantum Gauss and quantum diffeomorphisms constraints.  On the other
hand,
each element of ${\cal H}^G_{\rm Diff}$ is a linear functional defined on 
${\cal D}_{\rm kin}^G$,   
\begin{equation}
{\cal H}^G_{\rm Diff}\ \subset\ \left({\cal D}_{\rm kin}^G\right)^*  
\end{equation}
where the right hand side is the space of the linear functionals ${\cal D}_{\rm
kin}^G\rightarrow \mathbb{C}$.  We still use that extra structure intensively.
In particular, an operator 
$$\hat{{\cal O}}:{\cal D}_{\rm kin}^G\rightarrow {\cal D}_{\rm kin}^G$$ 
will be pulled back to the dual operator
$$\hat{{\cal O}}^*:{\cal D}^G_{\rm Diff}\rightarrow ({\cal D}_{\rm kin}^G)^*.$$

\subsection{The scalar constraint} 
In the Hilbert space ${\cal H}^G_{\rm Diff}$ of solutions of the quantum Gauss
and vector constraint, we impose the quantum scalar constraint
\begin{equation}\label{constraint}
(\hat{\pi}(x)^*\ -\ \hat{h}(x))\Psi\ =\ 0.
\end{equation}
In \cite{DGKL,HannoKristina} it is argued that a general solution can be derived
if one is able to introduce an operator 
$$\exp i\int d^3x \hat{\varphi}(x)\hat{h}(x) : {\cal H}^G_{\rm Diff}\rightarrow
{\cal H}^G_{\rm Diff}$$
of suitable, but quite natural, properties.  
We will define now such an operator in the very space ${\cal H}^G_{\rm Diff}$
and see that it does have the desired properties. 

\subsubsection{Extra structure needed for $\hat{h}(x)$}
To deal with the operator  (distribution) $\hat{h}(x)$ we will need more
structure. 
For each graph $\gamma$ its set of nodes will be denoted by $\gamma^0$.
For every of the subspaces ${\cal D}_X\otimes {\cal D}_\gamma'{}^G$
(modulo the diffeomorphisms) it is convenient to consider the subgroup 
$${\rm Diff}_{X\cup \gamma^0}$$ 
of Diff set by the diffeomorphisms  which act as identity on the set $X$ as well
as on the set $\gamma^0$ of the nodes of $\gamma$.
We  repeat the construction of the averaging for the diffeomorphisms
Diff$_{X\cup \gamma^0}$,
\begin{equation}
{\cal D}_X\otimes {\cal D}_\gamma'{}^G\ni\psi\ \mapsto\ \tilde{\eta}(\psi)\ =\
\frac{1}{\tilde{n}_{X,\gamma}}\sum_{[\phi]\in {\rm Diff_{X\cup\gamma^0}}/{\rm
TDiff}_{X,\gamma}}\langle U(\phi)\psi|
\end{equation} 
where the number ${\tilde{n}_{X,\gamma}}$ will be fixed later to be consistent
with another map $\check{\eta}$ introduced below. For example, if $\psi\in {\cal
D}_X\otimes {\cal D}_\gamma'{}^G$  
is a simple tensor product
$$ \psi\ =\ \langle \pi|\otimes f_\gamma$$ 
then,  
\begin{equation}
\tilde{\eta}(\psi)\ =\ \frac{\langle
\pi|}{\tilde{n}_{X,\gamma}}\otimes\sum_{[\phi]\in {\rm
Diff_{X\cup\gamma^0}}/{\rm TDiff}_{X,\gamma}}\langle U(\phi)f_\gamma|.
\end{equation} 
     
Given a finite set $Y\subset\Sigma$, we consider all the spaces  ${\cal
D}_X\otimes {\cal D}_{\gamma}$ such that 
$$    X\cup \gamma^0\ =\ Y, $$
combine them into the space
$$\bigoplus_{(X,\gamma)}{\cal D}_X\otimes {\cal D}_{\gamma}'{}^G, $$
and combine the maps $\tilde{\eta}$ to a linear map 
$$ \tilde{\eta}: \bigoplus_{(X,\gamma)}{\cal D}_X\otimes {\cal D}_{\gamma}'{}^G\
\rightarrow\ \left({\cal D}_{\rm kin}'^G\right)^*, $$
and endow the image of this map 
$${\cal D}_{{\rm Diff}_{Y}}^G\ :=\ \tilde{\eta}\left(\bigoplus_{(X,\gamma)}{\cal
D}_X\otimes {\cal D}_{\gamma}'{}^G\right) , $$
with a scalar product  
$$(\tilde{\eta}(\psi_1)|\tilde{\eta(\psi_2)})_{\rm Diff_{Y}}\ :=\
[\tilde{\eta}(\psi_1)](\psi_2). $$
In this way we obtain  the  Hilbert space 
$$ {\cal H}_{{\rm Diff}_{Y}}^G\ =\ \overline{{\cal D}_{{\rm Diff}_{Y}}^G},$$
that is needed to deal with the $\hat{h}(x)$ operator. 

The original averaging map $\eta$ for $\psi\in{\cal D}_X\otimes{\cal
D}_\gamma^G$
can be written as averaged  $\tilde{\eta}$,
$$ \eta(\psi)\ =\ \frac{1}{|Y|!}\sum_{[\phi]\in {\rm Diff}/{\rm Diff}_{Y}}
U(\phi)^* \tilde{\eta}(\psi), $$
where the choice of the normalization factor as the number of the
elements of the symmetry group of the set $Y$ is the condition that fixes the 
number $\tilde{n}_{X,Y}$ uniquely. 
The map  $\tilde{\eta}(\psi)\mapsto \eta(\psi)$ extends by the continuity  to
\begin{equation}\label{checketa}
{\cal H}_{{\rm Diff}_{Y}}^G\ \rightarrow\ {\cal H}_{{\rm Diff}}^G,
\ \ \ \ \ \check{\eta}(\tilde{\Psi})\ =\ \frac{1}{|Y|!}\sum_{[\phi]\in {\rm
Diff}/{\rm Diff}_{Y}} U(\phi)^*\tilde{\Psi}.
\end{equation}  
The factor $|Y|!$ ensures, that for every  $\tilde{\Psi}_I$, $I=1,2$ invariant
with respect to all Diff$_Y$,
$$ (\check{\eta}(\tilde{\Psi}_I)|\check{\eta}(\tilde{\Psi}_J))_{\rm Diff}\ =\
(\tilde{\Psi}_I|\tilde{\Psi}_J)_{{\rm Diff}_Y}.$$

Before we apply this structure to the operator $\hat{h}(x)$, let us use it to
characterize the action of the operator $\hat{\pi}(x)^*$ 
defined by the duality on the diffeomorphism invariant states, elements of the
space ${\cal D}_{\rm diff}^G \subset ({\cal D}_{\rm kin}^G)^*$. 
 It will be convenient to introduce for each $y\in \Sigma$, an operator
$\hat{\pi}_y$ defined in (a suitable domain of) ${\cal H}_{\rm kin,mat}$ by
$\hat{\pi}(x)$, 
\begin{equation}\label{pi(x)}
\hat{\pi}(x)\ =\ \sum_{y\in\Sigma}\delta(x,y)\hat{\pi}_y,\ \ \ \ \ \ \ \ 
\hat{\pi}_y|\pi\rangle\ =\ \pi_y|\pi\rangle \ , 
\end{equation}  
(recall that $\pi_y$ is not zero only for a finite set of points $y$). 
This definition passes by the duality to the  (bra) states 
\begin{equation}\label{pi_x}
\langle \pi|\hat{\pi}_y\ =\ \pi_y\langle \pi|. \ 
\end{equation}
Next, increasing the level of complexity, consider the action of the  operator
$\hat{\pi}_y^*$ in
each of the spaces  $ {\cal D}_{{\rm Diff}_{Y}}^G $. To begin with
$$ y\notin Y\ \Rightarrow\ \hat{\pi}_y^*|_{{\cal D}_{{\rm Diff}_{Y}}^G }\ =\ 0.
$$
The elements $\tilde{\eta}(\langle \pi|\otimes f_\gamma)$ are eigenvectors,
$$  \hat{\pi}_y^*\tilde{\eta}(\langle \pi|\otimes f_\gamma) \ =\ \pi_y 
\tilde{\eta}(\langle \pi|\otimes f_\gamma).  $$
Finally, to write the action of $\hat{pi}_y^*$ in ${\cal H}_{\rm Diff}^G$, 
given  
$$\tilde{\Psi}\ \in\ {\cal D}_{\rm Diff_Y}^G, \ \ \ \ {\rm and}\ \ \ \
\check{\eta}(\tilde{\Psi})\ \in\ {\cal H}_{\rm Diff}^G$$
we have
\begin{equation}\label{piY} \hat{\pi}_y^*\check{\eta}(\tilde{\Psi})\ =\
\frac{1}{|Y|!}\sum_{y'\in Y}\sum_{[\phi_{y'}]}
\hat{\pi}_y^*U({\phi_{y'}})^*\tilde{\psi}\ =\  \frac{1}{|Y|!}\sum_{y'\in
Y}\sum_{[\phi_{y'}]} U({\phi_{y'}})^*\hat{\pi}_{y'}^*\tilde{\psi}\end{equation} 
where for every $y'\in Y$, $[\phi_{y'}]$ runs through the subset of 
${\rm Diff}/{\rm Diff}_{Y}$  such that
$$ \phi_{y'}(y)\ =\ y'. $$
For 
$$\tilde{\Psi}\ =\ \langle\pi|\otimes \tilde{f}, $$
we have
$$ \hat{\pi}_y^*\check{\eta}(\langle\pi|\otimes \tilde{f})\ =\ 
\frac{1}{|Y|!}\sum_{y'\in Y} \pi_{y'}\sum_{[\phi_{y'}]}
U({\phi_{y'}})^*\langle\pi|\otimes \tilde{f}\ .$$   

The result of the action is not any longer an element of ${\cal H}_{\rm
diff}^G$, however the operator $\hat{\pi}_y$ is well defined
in the domain  ${\cal D}_{\rm diff}^G\subset {\cal H}_{\rm diff}^G$ in the
following sense
\begin{equation}
\hat{\pi}_y:\ {\cal D}_{\rm diff}^G\ \rightarrow\ ({\cal D}_{\rm kin}^G)^*\ .
\end{equation} 

Now, we are in the position to write down the action of the operator
$\hat{h}(x)$
 apparent in the quantum scalar constraint. It is not defined  directly in
${\cal H}_{\rm kin,gr}$, however it is defined in the spaces ${\cal H}_{{\rm
Diff}_Y}^G$. Actually, it is introduced in the opposite order
\cite{DGKL,HannoKristina} then the calculation of the action of $\hat{\pi}(x)$
was performed above. 
  
First, in each of the spaces ${\cal H}_{{\rm Diff}_{Y}}^G$ and for every $y\in
\Sigma$ the operator $\hat{h}_y$ is defined as a self-adjoint operator. The
operator is identically zero unless $y\in Y$, 
$$ y\notin Y\ \Rightarrow\ \hat{h}_y|_{{\cal H}_{{\rm Diff}_{Y}}^G}\ =\ 0. $$

By the linearity, $\hat{h}_y$ is extended to the span
\begin{equation}\label{theSpan} 
{\rm Span}\left( {\cal H}_{{\rm Diff}_{Y}}^G\ :\ Y\subset \Sigma, \ \ |Y|<\infty
\right)\ \subset\ \left({\cal D}_{\rm kin}^G\right)^*. 
\end{equation}
For different points the operators commute,
\begin{equation}
y\not=y'\ \Rightarrow\ [\hat{h}_{y},h_{y'}]\ =\ 0  .
\end{equation}

The map $y\mapsto \hat{h}_y$ is diffeomorphism invariant in the sense that
for every diffeomorphism $\phi\in$Diff and its (dual) action $U(\phi)^*$ in the
subset (\ref{theSpan}) of $({\cal D}_{\rm kin}^G)^*$ we have 
$$ \hat{h}_{\phi^{-1}(y)}U(\phi)^*\  =\  U(\phi)^*\hat{h}_y. $$

The action of $\hat{h}_y$ is ${\cal H}_{\rm Diff}^G$ is defined by the analogy
to (\ref{piY}), that is given 
$$\tilde{\Psi}\ \in\ {\cal D}_{\rm Diff_Y}^G, \ \ \ \ {\rm and}\ \ \ \
\check{\eta}(\tilde{\Psi})\ \in\ {\cal H}_{\rm Diff}^G$$
we have
$$ \hat{h}_y\check{\eta}(\tilde{\Psi})\ =\ \frac{1}{|Y|!}\sum_{y'\in
Y}\sum_{[\phi_{y'}]} \hat{h}_yU({\phi_{y'}})^*\tilde{\psi}\ =\ 
\frac{1}{|Y|!}\sum_{y'\in Y}\sum_{[\phi_{y'}]}
U({\phi_{y'}})^*\hat{h}_{y'}\tilde{\psi}$$ 
where the notation is the same as in (\ref{piY})

\subsubsection{The exp$(\int d^3x \hat{\varphi}(x)\hat{h}(x))$ operator}
We can turn now, to the introduction of  an operator exp$(i\int
d^3x\hat{\varphi}(x)\hat{h}(x))$. For every of the spaces ${\cal H}_{{\rm
Diff}_{Y}}^G$ there is a basis of simultaneous eigenvectors of the operators
$\hat{h}_{y}$ and $\hat{\pi}_y$, $y\in\Sigma$. We choose a one, and denote its
elements by $\langle \pi|\otimes\langle h,\alpha|$ where   
$$ h:  y\mapsto h_y,\ \ \ \ \ \ \pi: y\mapsto \pi_y$$
are functions  of finite supports such that    
\begin{equation}
\hat{h}_{y}\langle \pi|\otimes\langle h,\alpha|\ =\ {h}_{y}\langle
\pi|\otimes\langle h,\alpha|, \ \ \ \hat{\pi}_{y}\langle \pi|\otimes\langle
h,\alpha|\ =\ \pi_{y}\langle \pi|\otimes\langle h,\alpha|
\end{equation}          
and $\alpha$ is an extra label.
We define (compare with (\ref{exphpi}))
$$e^{i\int d^3x\hat{\varphi}(x)\hat{h}(x)}\langle \pi|\otimes\langle h,\alpha|\
=\ \langle \pi+h|\otimes\langle h,\alpha|.$$     
That defines an operator in each of the spaces ${\cal H}_{\rm Diff_{Y}}$
and in the span which is the direct (orthogonal) sum (\ref{theSpan})

This operator is unitary, 
\begin{equation}\label{expiphihdagger} \left(e^{i\int
d^3x\hat{\varphi}(x)\hat{h}(x)}\right)^\dagger\ =\ 
e^{-i\int d^3x\hat{\varphi}(x)\hat{h}(x)}, \end{equation}
where the right hand side is defined by 
$$e^{-i\int d^3x\hat{\varphi}(x)\hat{h}(x)}\langle \pi|\otimes\langle h,\alpha|\
=\ \langle \pi-h|\otimes\langle h,\alpha|. $$     
The operator is diffeomorphisms invariant,
\begin{equation}
U(\phi)^* e^{i\int d^3x \hat{\varphi(x)}\hat{h}(x)}\ =\   e^{i\int
d^3x\hat{\varphi(x)}\hat{h}(x)}U(\phi)^*.  
\end{equation}

Finally, to define this operator in ${\cal H}_{\rm Diff}^G$, for every
$\tilde{\Psi} \in {\cal H}_{\rm Diff_{Y}}$ and the corresponding
$\check{\eta}(\tilde{\Psi})\in {\cal H}_{\rm Diff}^G$ we write 
\begin{equation}
e^{i\int d^3x\hat{\varphi}(x)\hat{h}(x)} \check{\eta}(\tilde{\Psi})\ :=\ 
 \frac{1}{|Y|}\sum_{[\phi]\in {\rm Diff}/{\rm Diff}_{Y}} e^{i\int
d^3x\hat{\varphi}(x)\hat{h}(x)}U(\phi)^*\tilde{\Psi}.
\end{equation}
Indeed, we can always do it, but is the right hand side again an element of the
Hilbert space ${\cal H}_{\rm Diff}^G$? The answer is affirmative due to the
diffeomorphism invariance, namely, it follows that 
\begin{equation}\label{expfinal}
e^{i\int d^3x\hat{\varphi}(x)\hat{h}(x)} \check{\eta}(\tilde{\Psi})\ =\ 
\check{\eta}(e^{i\int d^3x\hat{\varphi}(x)\hat{h}(x)} \tilde{\Psi})\ \in\ 
{\cal H}_{\rm Diff}^G .
\end{equation}
The extension by the linearity and continuity provides a unitary operator
$$  e^{i\int d^3x\hat{\varphi}(x)\hat{h}(x)}: {\cal H}_{\rm Diff}^G\rightarrow 
{\cal H}_{\rm Diff}^G $$
for which the property (\ref{expiphihdagger}) still holds. 

Now, it is not hard to check, that our operator (\ref{expfinal}) does satisfy
the desired property, namely for  every $\Psi\in {\cal H}_{\rm Diff}^G$, 
\begin{equation}
e^{-i\int d^3x\hat{\varphi}(x)\hat{h}(x)}\left(
\hat{\pi}(y) - \hat{h}(y) \right)e^{i\int d^3x\hat{\varphi}(x)\hat{h}(x)} \Psi \
=\ 
\hat{\pi}(y) \Psi\ \in \ \left({\cal D}_{\rm Diff}^G\right)^* .
\end{equation}
\subsubsection{Solutions, Dirac observables, dynamics}
The quantum scalar constraint  
\begin{equation}\label{qscalarconstraint} 
(\hat{\pi}(x) - \hat{h}(x)) \Psi\ =\ 0 
\end{equation}
is equivalent to  
$$ \hat{\pi}(x) e^{-i\int d^3x\hat{\varphi}(x)\hat{h}(x)}\Psi\ =\ 0 .$$
Moreover, the condition on the Dirac observable 
$$ [\hat{\pi}(x) - \hat{h}(x),\hat{\cal O}]\ =\ 0$$
is equivalent to
$$ [\hat{\pi}(x),\ e^{-i\int d^3x\hat{\varphi}(x)\hat{h}(x)}\hat{\cal O}e^{i\int
d^3x\hat{\varphi}(x)\hat{h}(x)}]\ =\ 0\ .$$
In ${\cal H}_{\rm Diff}^G$, solutions to the equation
$$\hat{\pi}(x)\Psi'\ =\ 0 $$ 
set the the subspace given by
$$ \overline{\eta\left( |0\rangle\otimes \bigoplus_{\gamma}{\cal
D}'_\gamma{}^G\right)}\ =\ {\cal H}_{\rm Diff,gr}^G,$$ 
that is the subspace of states independent of $\varphi$. Hence,  
solutions to the quantum scalar (and the Gauss) constraint are     

\begin{equation}
{\cal H}_{\rm Diff}^G\ \ni \Psi\ =\ e^{i\int
d^3x\hat{\varphi}(x)\hat{h}(x)}\Psi',\ \ \ \ \ \ \ \ \Psi'\in {\cal H}_{\rm
Diff,gr}^G.
\end{equation}
Denote the subspace they set by
\begin{equation}
{\cal H}_{\rm phys}\ \subset\ {\cal H}_{\rm Diff}^G. 
\end{equation}
A Dirac observable is every operator  
\begin{equation}
e^{i\int d^3x\hat{\varphi}(x)\hat{h}(x)}\hat{L}e^{-i\int
d^3x\hat{\varphi}(x)\hat{h}(x)},
\end{equation}
defined  in ${\cal H}_{\rm phys}$ by an operator $\hat{L}$ defined in
${\cal H}_{\rm Diff,gr}^G$.  
Another observable can be defined from the operators $\hat{\pi}(x)$, 
for example
$$\int d^3 x\hat{\pi}(x)$$
however,
$$ \hat{\pi}(x)|_{{\cal H}_{\rm phys}}\ =\ \hat{h}(x)|_{{\cal H}_{\rm phys}} $$
and $\hat{h}(x)$ is defined in ${\cal H}_{\rm Diff,gr}^G$.
Our map (\ref{expfinal}) can be generalized to a family of maps corresponding to
the transformation  $\phi\mapsto \phi+\tau$, $\tau\in\mathbb{R}$. For every
$\tau$ the transformation should amount to a transformation  
\begin{equation}
e^{i\int d^3x \tau\hat{h}(x)} : {\cal H}_{\rm Diff}^G\rightarrow {\cal H}_{\rm
Diff}^G, 
\end{equation} 
where the operator has to be defined. To define the  operator exp$({i\int d^3x
\tau \hat{h}(x)})$ we repeat the construction that lead us to the operator  
exp$({i\int d^3x \hat{h}(x)})$, with the starting point
$$e^{i\int d^3x\tau\hat{h}(x)}\langle \pi|\otimes\langle h,X,\gamma^0,\alpha|\ =\
e^{i\sum_x h_x}\langle \pi|\otimes\langle h,X,\gamma^0,\alpha| .$$     
As expected, the operator preserves the space of solutions
$$e^{i\int d^3x \tau\hat{h}(x)} \left( {\cal H}_{\rm phys} \right)\ =\ 
 {\cal H}_{\rm phys} $$  
and defines therein the dynamics.

\section{Summary and seeds of a new idea}

The first conclusion is that a quantization of the scalar field whose existence
and suitable properties were assumed in \cite{DGKL} exists, and an example is
the polymer quantization. Furthermore, it is shown explicitly, that  as argued
in \cite{DGKL}, the theory is equivalent to the quantum theory in the Hilbert
space ${\cal H}_{\rm Diff,gr}^G$ of diffeomorphism invariant states of the
gravitational degrees of freedom only, with the dynamics defined by the physical
Hamiltonian
\begin{equation}
\hat{h}_{\rm phys}\ =\ \int d^3x \hat{h}(x),
\end{equation}  
where $\hat{h}(x)$ is a quantization of the classical solution for $\pi(x)$  
$$ \pi(x)\ =\ h(x) $$  
following from the constraints. In this way, the current work completes the
derivation of the model already formulated in \cite{DGKL}.   
Technically, we have implemented in detail the diffeomorphism averaging for loop
quantum gravity states of geometry coupled with the polymer states of scalar
field and discussed the general structure of the operators emerging in the
scalar constraints.  Mathematically, the physically relevant part of the Hilbert
space ${\cal H}_{\rm Diff}^G$ is contained in the so called habitat space
introduced in \cite{LewandMarolf}. This is a new application of the habitat
framework which may be useful for various technical questions.  

Secondly, it turns out, that in the framework of the polymer quantization of the
scalar field, the Hilbert space ${\cal H}_{\rm phys}$ of the physical states,
solutions to the quantum constraints,  is a subspace   
of the Hilbert space of solutions to the diffeomorphism constraint,
$$ {\cal H}_{\rm phys}\ \subset {\cal H}_{\rm Diff}^G.$$   
Therefore, more structure is at our disposal, than only the physical states
themselves. This  advantage is not only estetic. It also gives a clue for quite
promising development of the theory. We explain this below.    

The classical constraints  for the massless field coupled to gravity
are 
\begin{align} C(x)\ &= C^{\rm gr}(x)\ +\ \frac{1}{2}\frac{\pi^2(x)}{\sqrt{q(x)}}
+\frac{1}{2}q^{ab}(x)\phi_{,a}(x)\phi_{,b}(x)\sqrt{q(x)}, \label{scalar}\\
C_a(x)\ &=\  C^{\rm gr}_a(x) \ +\ \pi(x)\phi_{,a}(x).\label{vector}
\end{align}
where $q_{ab}$ is the 3-metric tensor induced on a 3-slice of spacetime
$C^{\rm gr}$ is the gravitational field part of the scalar constraint, and 
$C^{\rm gr}_a$ is the gravitational part of the vector constraint.

The scalar constraint $C(x)$ can be replaced by $C'(x)$ (deparametrized scalar
constraint):
\begin{align}
C'(x)\ &=\ \pi^2(x) - h^2(x),\\
h_{\pm}\ &:=\ \sqrt{-\sqrt{q}C^{\rm gr }+/- 
\sqrt{q}\sqrt{({C^{\rm gr}})^2-q^{ab}C^{\rm gr}_a C^{\rm gr}_b}}.\label{hclas}
\end{align}

The sign $\pm$ in $h_{\pm}$ is $+$ in the part of the phase space 
at which 
\begin{equation} \pi^2 \ \ge\  \phi_{,a}\phi_{,b}q^{ab} \det q,\label{+} 
\end{equation}
for example in the neighborhood of the homogeneous solutions.  

The sign $\pm$ in $h_{\pm}$ is $-$, on the other hand, in the part of the phase
space
at which
\begin{equation} \pi^2 \ \le\  \phi_{,a}\phi_{,b}q^{ab} \det q. \label{-}
\end{equation}

Each of the cases (\ref{+},\ref{-}) consists of two in cases,
\begin{equation}
\pi(x) \ =\ +h_\pm(x), \ \ \ \ \ {\rm or}\ \ \ \ \ \ \pi(x)\ =\
-h_\pm(x).\label{pmpm} 
\end{equation} 

A natural first goal \cite{DGKL}, was to restrict the quantization to the  case
(\ref{+}) and positive $\pi$,  and quantize the theory for the part of the phase
space which contains expanding homogeneous solutions. Now, the formulation of
the current paper allows an attempt to unify the theory to the both  cases
(\ref{+}) and (\ref{-}) the both cases (\ref{pmpm}). Indeed, we can accommodate
in the Hilbert space ${\cal H}_{\rm Diff}^G$ simultaneously  quantum solutions
to
each of the cases. To this end,  one has to implement the construction presented
in the current paper       for each of the following 4 cases
$$\hat{h}(x)\ =\ \hat{h}_+(x), -\hat{h}_+(x),\hat{h}_-(x), -\hat{h}_-(x).$$ 
The result will be four subspaces 
$$ {\cal H}_{{\rm phys}++},\ {\cal H}_{{\rm phys}-+},\ {\cal H}_{{\rm
phys}+-},\ {\cal H}_{{\rm phys}--}\ \subset\ {\cal H}_{\rm Diff}^G. $$
They span the total space of solutions
$${\cal H}_{\rm phys}\ =\ {\rm Span}({\cal H}_{{\rm phys}++},\ {\cal H}_{{\rm
phys}-+},\ {\cal H}_{{\rm phys}+-},\ {\cal H}_{{\rm phys}--})\ 
\subset\ {\cal H}_{\rm Diff}^G . $$
The space is  endowed with the evolution induced by the transformation 
$$ \varphi\ \mapsto\ \varphi+\tau .$$
Whether this is it, or more input is needed is an open question. In any case. 
certainly, this framework takes us beyond the state of art.

\section{Acknowledgements}

We benefited a lot from comments of Andrea Dapor, Pawe{\l} Duch and Hanno
Sahlmann.  This work was partially supported by the grant of Polish Ministerstwo
Nauki i Szkolnictwa Wy\.zszego
nr  N N202 104838 and by the grant of Polish Narodowe Centrum Nauki nr
2011/02/A/ST2/00300.

\end{document}